\documentclass[12pt,preprint]{aastex}
\usepackage{color}

\newcommand{\nv}{N\,{\footnotesize V}}
\newcommand{\civ}{C\,{\footnotesize IV}}
\newcommand{\siiv}{Si\,{\footnotesize IV}}
\newcommand{\mgii}{Mg\,{\footnotesize II}}
\newcommand{\caiik}{Ca\,{\footnotesize II} K}
\newcommand{\naid}{Na\,{\footnotesize I} D}

\newcommand{\aliii}{Al\,{\footnotesize III}}

\newcommand{\oii}{[O\,{\footnotesize II}]}

\newcommand{\neiii}{[Ne\,{\footnotesize III}]}

\newcommand{\hb}{H$\beta$}

\newcommand{\hei}{He\,{\footnotesize I*}}

\newcommand{\heiteen} {He\,{\footnotesize I*} $\lambda$3889}
\newcommand{\heiozetz}{He\,{\footnotesize I*} $\lambda$10830}

\newcommand{\pag} {Pa$\gamma $}

\newcommand{\oiii}{[O\,{\footnotesize III}]}

\newcommand{\feii}{Fe\,{\footnotesize II}}

\newcommand{\kmps}{$\rm km~s^{-1}$}

\newcommand{\ergs}{$\rm erg~s^{-1}$}

\begin{document}
\title{\bf A Broad Absorption Line Outflow Associated with the Broad Emission Line Region in the Quasar SDSS J075133.35+134548.3}
\author{Bo Liu\altaffilmark{1,2}, Hongyan Zhou\altaffilmark{1,2}, Xinwen Shu\altaffilmark{3}, Shaohua Zhang\altaffilmark{2},Tuo Ji\altaffilmark{2}, Xiang Pan\altaffilmark{2},Peng Jiang\altaffilmark{2}}
\altaffiltext{1}{Key Laboratory for Research in Galaxies and Cosmology, University of Science and Technology of China, 96 Jinzhai Road, Hefei, Anhui, 230026}
\altaffiltext{2}{Polar Research Institute of China, 451 Jinqiao Road, Shanghai, 200136, China}
\altaffiltext{3}{Department of Physics, Anhui Normal University, Wuhu, Anhui, 241000, People’s Republic of China; xwshu@mail.ahnu.edu.cn 0000-0002-7020-4290}
\begin{abstract}
We report on the discovery of unusual broad absorption lines (BALs) in the bright quasar SDSS J075133.35+134548.3 at z $\sim$ 1, using archival and newly obtained optical and NIR spectroscopic data. The BALs are detected reliably  in \heiteen, \heiozetz\ and tentatively in \aliii, \mgii.  These BALs show complex velocity structures consisting of two major components: a high-Velocity component (HV), with a blueshifted velocity range of $\Delta v_{\rm HV}$ $\sim$ -9300 -- -3500 \kmps, can be reliably detected in  \heiozetz, and tentatively in \aliii\ and \mgii, whereas it is undetectable in \heiteen; and a low-Velocity component (LV), with $\Delta v_{\rm LV}$$\sim$-3500 -- -1800 \kmps, is only detected in \heiteen\ and \heiozetz. With the BALs from different ions, the HV outflowing gas can be constrained to have a density of $n_{\rm H}$$\sim$10$^{10.3}$--10$^{11.4 }$ cm$^{-3}$, a column density of $N_{\rm H}$ $\sim$ 10$^{21} $cm$^{-2}$ and an ionization parameter of $U$ $\sim$10$^{-1.83}$--10$^{-1.72 }$;  inferring a distance of $R_{\rm HV}$$\sim$0.5 pc from the central continuum source with a monochromatic luminosity $\lambda L_{\lambda}$(5100)=7.0$\times$10$^{45}$ \ergs\ at 5100 \AA. This distance is remarkably similar to that of the normal broad emission line region (BLR) estimated from the reverberation experiments, suggesting associationof the BLR and the HV BAL outflowing gas. Interestingly, a blueshifted component is also detected in \aliii\ and \mgii\ broad emission lines (BELs), and the \aliii/\mgii\ of such a BEL component can be reproduced by the physical parameters inferred from the HV BAL gas. The LV BAL gas likely has a larger column density, a higher ionization level, and hence a smaller distance than the HV BAL gas. Further spectroscopy with a high S/N ratio and broader wavelength coverage is needed to confirm this to shed new light on the possible connection between BALs and BELs.

\end{abstract}
\section{Introduction}
As a regulator of the growth of central supermassive black holes (BH) in active galactic nuclei (AGNs), outflows carry away the angular momentum of inflowing gas to sustain mass accretion.  (Sulentic et al. 2000; Richards et al. 2011; Wang et al. 2011). Previous studies (e.g. Silk \& Rees 1998), suggested that outflows in AGNs can change the gas distribution  and hence influence star formation rates in the host galaxies.
Blueshifted broad emission lines (BELs; Gaskell 1982) and broad
absorption lines (BALs; Weymann et al. 1991) in the quasar spectra are the
prominent imprints of outflows\textbf, from which the physical condidtions of AGNs and their surounding can be
studied in detail.

Traditional BALs defined by BI (Weymann et al.  1991) in quasar spectra are often blueshifted at high speeds (up to about 0.2c) and
significantly broadened ($\geq$ 2000 \kmps). These BALs are from high ionization ions, such as \nv, \civ\ and \siiv, and low ionization ions, such as \aliii\ and \mgii\ (Hall et al. 2002;  Hewett \& Foltz 2003; Trump et al. 2006;  Zhang et al. 2010, 2014, 2017).
 Because of their prominent features, most of the BALs  are easily detected and measured.  Making use of BALs from different ions, we can set constraints on the physical conditions of the outflows.
However, the global  geometry, for example the covering factor, cannot be determined through the BALs,  due to only a single line of sight. Conversely, the global covering fraction of the outflows can be constrained by  blueshifted BELs.
Unfortunately, the blueshifted BELs are often not isolated and its decomposition from the the normal BELs of the broad line region (BLR), is challenging. In fact, the existence of the blueshifted BELs  is usually identified  through their different profiles, such as asymmetry , peak and/or line centroid, compared to the normal BELs  (e.g., Gaskell 1982; Crenshaw 1986; Marziani et al. 1996; Richards et al. 2002; Boroson 2005; ; Crenshaw et al. 2010; Rafiee et al. 2016; Zhang et al. 2017; Liu et al. 2019).

 The co-existence of both blueshifted BELs and BALs have been reported in several quasars and their physical connections have been inferred. Liu et al. (2016) presented a detailed
study  of BEL and BAL  outflows in a quasar, namely SDSS J163459.82+204936.0 (hereafter J1634)
and found  that their physical parameters extracted from
the photoionization code CLOUDY are  similar, strongly suggesting that the observed blueshifted BELs are emitted from the BAL outflowing gas.
More recently, Liu et al. (2019)  reported another quasars SDSS
J163345.22+512748.4 (hereafter J1633) in  which  both blueshifted BELs and BALs  were detected.
The inferred physical conditions are also  similar, except for the column density,
 indicating that the blueshifted BEL outflow and BAL outflow may  be associated.
If the blueshifted BELs and BALs are indeed from the same outflowing gas, the joint
analysis of BELs and BALs could provide  complementary or additional constraints on the general physical
properties of outflows in quasars.

In this paper, we  present a detailed analysis of the BAL and BEL  outflows of the quasar
SDSS J075133.35+134548.3 (hereafter SDSS J0751+1345).
 By combining BALs of the \heiozetz, \mgii, and \aliii, the properties of  the BAL outflow
can be well determined. In particular, two BAL components with different outflowing
 velocities are revealed. In \aliii\ and \mgii\ BELs, we also  detect the presence of the blueshifted BEL components. Although the characters of the outflow gas, from which the blueshifted BELs   originate,  are not well constrained as  those of  the BAL outflow gas, our analysis suggests that they are not isolated.
We describe the observation data in Section 2 and the data is further analysed in Section 3 to 6. Our  discussion on the results is present in Section 7.
In this paper, the cosmological parameters  are adopted as $H_0=70$~km~s$^{-1}$~Mpc$^{-1}$, $\Omega_{\rm M}=0.3$, and
$\Omega_{\Lambda}=0.7$.

\section{Observation and Data Reduction}
The optical photometric data of SDSS J0751+1345 was taken by  the Sloan Digital Sky Survey (SDSS) on December 13, 2004.
The point-spread function magnitudes are $19.16\pm0.02$, $18.42\pm0.01$,
$18.00\pm0.01$, $17.85\pm0.01$ and $17.50\pm0.02$  in the SDSS ugriz bands
respectively.  The optical spectrum of SDSS J0751+1345 was observed by the Baryon Oscillation Spectroscopic Survey (BOSS; Dawson et al. 2013) of the SDSS third stage (Eisenstein et al. 2011) on January 8, 2011.  The spectrum we used was extracted from the BOSS Date Release 10 (DR10; Ahn et al. 2014) and its wavelength coverage is from 3600 \AA\ to 10500 \AA.

\begin{figure*}[htbp]
\centering
\epsscale{1}
\plotone{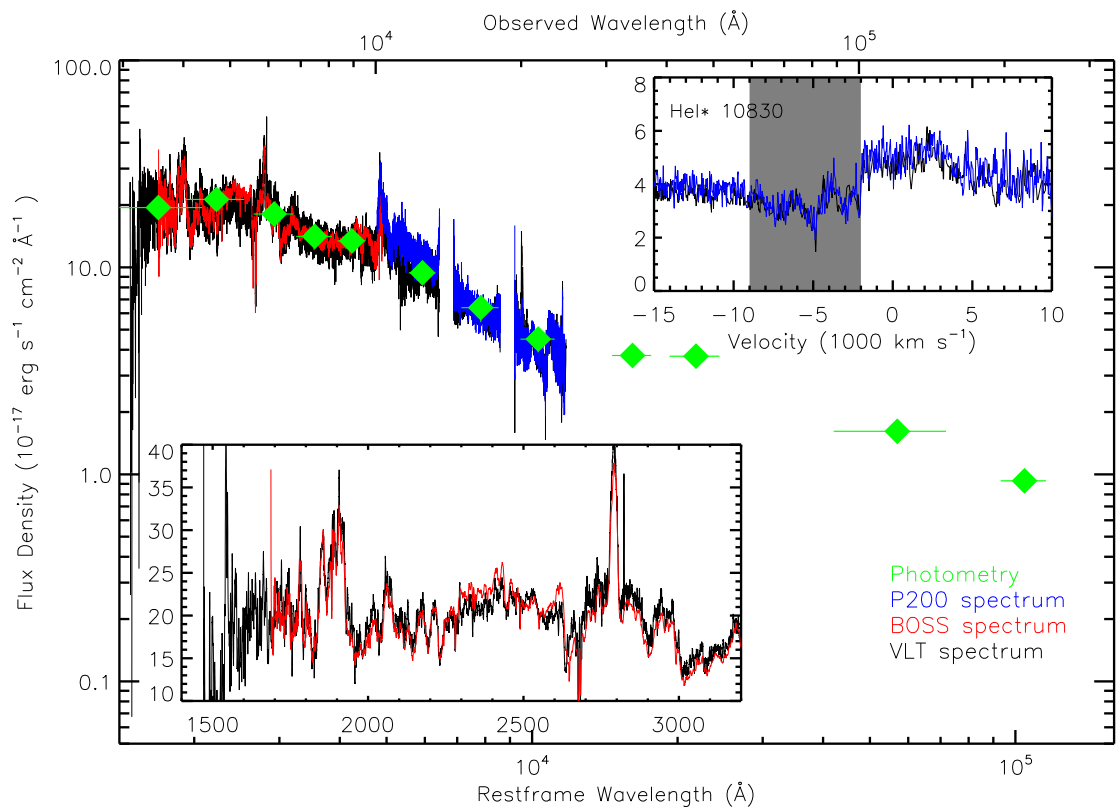}
\caption{ UV to mid-infrared spectra and SED of SDSS J0751+1345 in the
quasar's rest-frame. The BOSS, P200, and VLT spectra are  presented by red, blue and black curves, respectively.
The  photometric data is shown in green diamonds. The consistence between the spectral and photometric data
indicates that the variability between  the observations can be ignored and we can combine the spectra for the following analysis.
  The upper inset shows the detections of the \heiozetz\ BAL in K-band of P200 and VLT spectra. The  lower inset shows the BOSS and VLT spectral details between 1400 and 3200 \AA. }
\label{f1}
\end{figure*}

SDSS J0751+1345 was also observed  with the Very Large Telescope (VLT)/X-Shooter (Vernet et al. 2011) on Dec. 6, 2014 under  the ESO program 094.A-0087(A)\footnote{PI: Petitjean, Title: large XSHOOTER follow-up of peculiar BOSS quasars}. For the three arms, UVB, VIS and NIR, the exposure times are 2820 s, 2520 s and 2400 s , respectively. The slit widths of  the UVB, VIS and NIR arms are 1.6\arcsec ,1.5\arcsec and 1.2\arcsec,  leading to spectral  resolution ($R=\lambda/\delta\lambda$) of 1900, 3200 and 3900, respectively. Reduced 1-D spectra for  the three arms are retrieved from ESO Phase 3 Data Release, and  concatenated to form a single spectrum covering 3200 \AA --2.5 $\mu$m. Finally, telluric absorption features are corrected using Molecfit (Smette et al. 2015) and the final spectrum (hereafter the VLT spectrum) is shown in Fig.\ref{f1}.

Besides the optical band, SDSS J0751+1345  was also detected in the surveys of of the two micron all sky survey (2MASS; Skrutskie et al. 2006) and the Wide-field Infrared Survey Explorer (WISE; Wright et al. 2010). All the photometric data of SDSS J0751+1345  are listed in Table.1. On January 10, 2017,  using the TripleSpec  (Wilson et al. 2004)  on the 200 inch Hale telescope at Palomar Observatory, we  obtained its near infrared (NIR)  spectrum (hereafter P200 spectrum) in an A-B-B-A dithering mode. Four exposures of 150s  each were taken with the primary configuration of the instrument.  To match the seeing during the observation, we chose the width of the slit as 1\arcsec. The corresponding spectral resolution  was about 2700 and the wavelength coverage  was about  0.95--2.46 $\mu$m.   For the flux calibration,  we also observed two telluric standard stars quasi-simultaneously. Two gaps caused by the atmosphere transmissivity exit around 1.35 and 1.85 $\mu$m in the NIR spectrum.  According to the observed results displayed in Fig.\ref{f1}, SDSS J0751+1345 shows no obvious variability  in between the multi-spectrographic observations. Besides the global spectral feature, a HeI* 10830 BAL was also detected in K-band in both the P200 and VLT spectra and the profiles in the two spectra  were similar to each other. Thus, to increase the spectral signal to noise ratio (SNR), we created
a new spectrum from UV to NIR by combining the three spectra after masking the pixels which  were bad or polluted seriously by skylines.

\section{Systematic Redshift}
Before the spectral analysis, we first  attempted to derive the systematic redshift of SDSS J0751+1345. For  quasars, one of the main approaches to obtain their redshifts is through  to compare the  observed wavelengths of individual emission lines and their rest wavelengths (e.g., Paris et al. 2012).
For SDSS J0751+1345,  a broad \hb\ is detected in the spectrum and the peak of multi-Gaussian  BEL can be employed to obtain the systematic redshift (Bonning et al. 2007; Shen et al. 2011)

The redshift of SDSS J0751+1345 was first set as z=1.10052, which  was derived from the BOSS DR10 catalogue (Paris et al. 2014).  Adopting this redshift, we made a detail spectral fitting around \hb\ and found  that the peak of  the \hb\ BEL  was shifted by about 650 \kmps. According to  previous study
of \hb\ in a large sample of quasars (Shen et al. 2011), the mean peak offset of the
\hb\ broad line is 150 $ \pm $ 200 \kmps\ and the measurement result of SDSS J0751+1345
is beyond  the 2$ \sigma $ range. Thus, the redshift for SDSS J0751+1345 from the BOSS release
may be slightly  offset due to the complex of line profiles, and we revised the redshift of SDSS J0751+1345  from the peak of  the broad \hb\ as 1.10512$ \pm $ 0.00032.

\begin{figure*}[htbp]
\centering
\epsscale{1}
\plotone{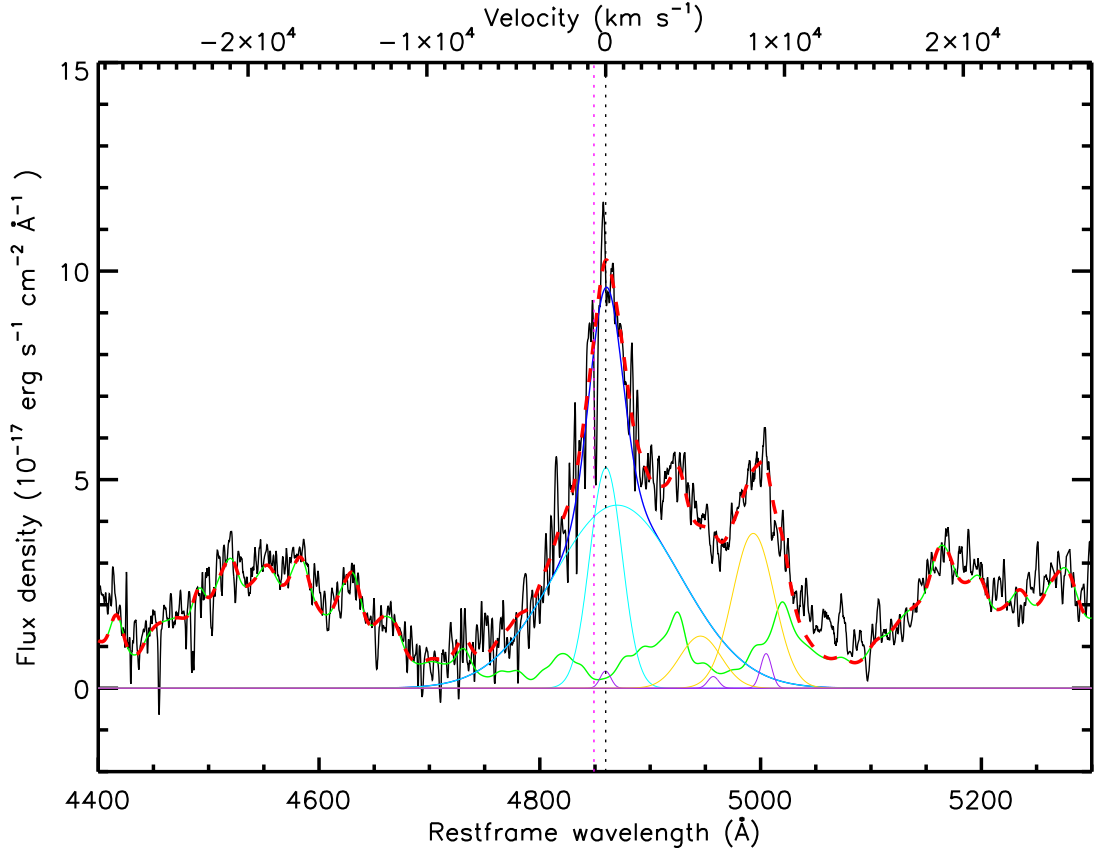}
\caption{Spectra of \hb\ for SDSS J0751+1345. The horizontal  axis is the rest-frame wavelength or relative velocity with redshift derived from the peak of broad \hb.
The black line shows the original continuum-subtracted spectrum,
while the  red dashed line shows the model including all emission-line components.
The blue line shows the total profile for the \hb\ emission and the cyan lines show the
individual Gaussian fittings.
The green lines represent optical \feii\ and the gold lines are the contributions associated
with the narrow-line region (\hb\ NEL, \oiii\ 4959,5007).
The red dotted line marks the zero-velocity according the redshift extracted from the BOSS DR10 catalogue
and it is about 650 \kmps\ blueshifted. }
\label{f2}
\end{figure*}

The profile of  the \hb\ BEL is derived as follows: first, a single power law continuum and  the optical \feii\ multiples  derived from I ZW1 (Veron-Cetty, Joly \& Veron 2004) are used to fit the spectrum in the wavelength range of 4000-5500\AA.  The fitting  procedure is  the same  as that described  in Dong et al. (2008).
After subtracting the continuum and optical \feii, we can derive the blend of  the \hb\ and \oiii\ emission lines.
The \hb\ BEL is modelled with one to four Gaussians. The fits are accepted when it cannot be improved significantly by  adding one more Gaussian (up to 4) with a chance probability  of less than 0.05 according to  the F-test. The fitting results indicate that, for SDSS J0751+1345, two Gaussians are good enough to fit the \hb\ BEL.  The \oiii\ doublet are fitted with the same profile and the line ratio \oiii\ 5007/\oiii\ 4959 is fixed 3. Each of them is modelled with two Gaussians for a core and a blue-shifted component, respectively (e.g. Komossa et al. 2008; Zhang et al. 2011). The \hb\ NEL is assumed to share the same profile with \oiii\ core.  The fitting results of  the emission lines, converted  to the revised redshift, are displayed in Fig.\ref{f2}. Also, based on
the profile of  the broad \hb, we derived the mass of  the central BH  to be log $M_{BH}/M_\odot=8.57\pm$ 0.37.

\begin{figure*}[htbp]
\centering
\epsscale{1}
\plotone{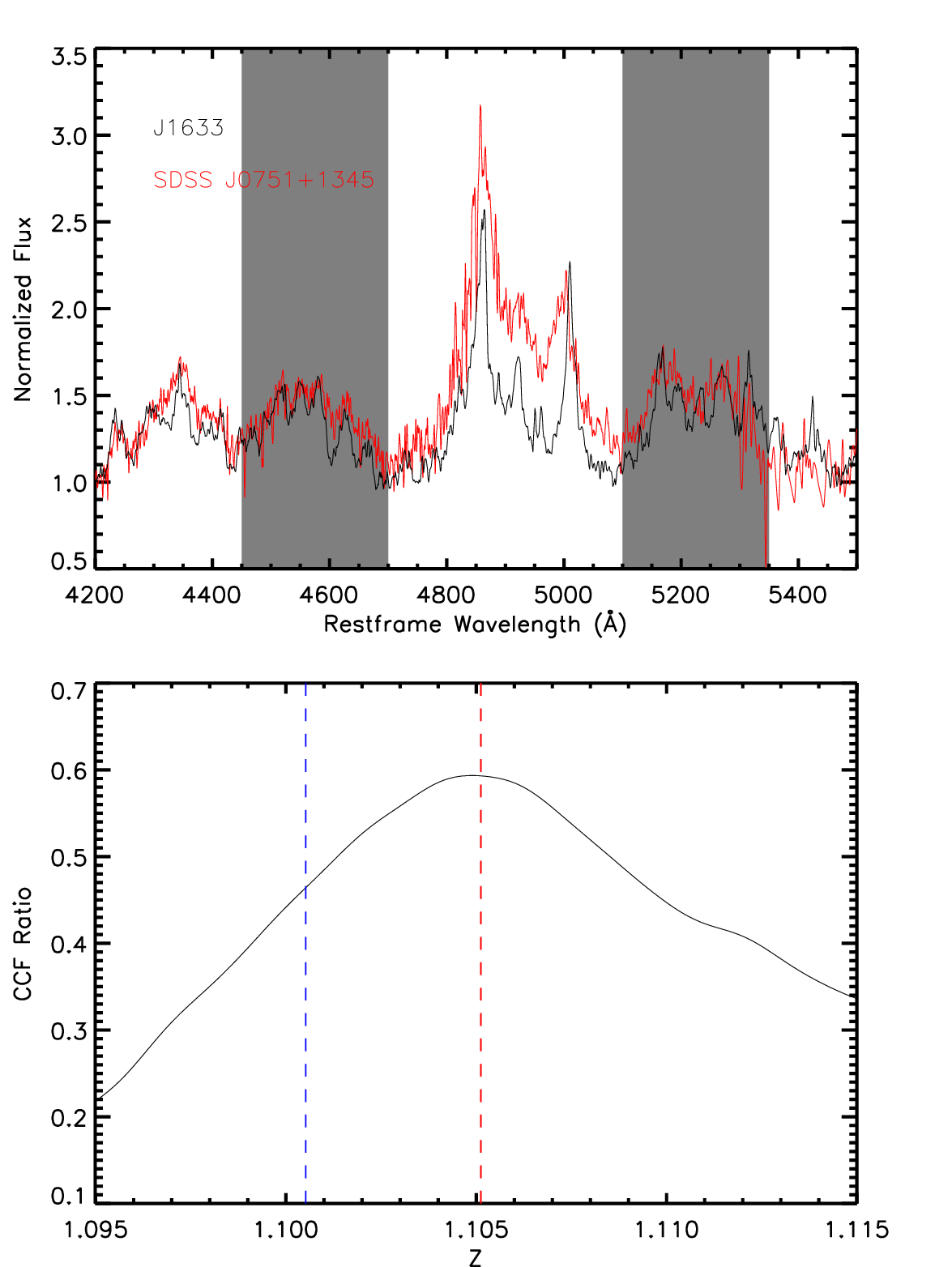}
\caption{\textit{Top}: Normalized spectrum of SDSS J0751+1345 (red) in the wavelength range of 4200--5500 \AA\ according to the revised redshift determined by broad \hb. The normalized spectrum of J1633 (black) is also displayed  for comparison. The spectral region of 4450--4700 \AA\ and 5100--5350 \AA, which is employed in the cross-correlation to verify the redshift revision, is marked in gray.
 \textit{Bottom}: Cross correlation between the  the spectrum of SDSS J0751+1345 and the corresponding region of J1633.  The redshifts of SDSS J0751+1345 obtained from  the DR10 catalogue and broad \hb\ are marked with blue and red dashed lines.  The peak of the curve indicates that the revised redshift derived from broad \hb\ is reliable.  }
\label{f3}
\end{figure*}

Besides the \hb\ BEL, the previous studies  suggested that the optical \feii\  multiples have no obvious offset  with respect to the intrinsic redshift (Sameshima et al. 2011). For SDSS J0751+1345, we also  attempted to use its optical \feii\ to verify the redshift revision.
As shown in the top panel of Fig.\ref{f3},  after converting to the quasar's rest-frame with the revised redshift, the spectrum of SDSS J0751+1345 is normalized with the continuum and  plotted in red.  For comparison,  the normalized spectrum of J1633 is displayed in red.  We  scaled the normalized J1633 spectrum with a constant to make sure the total flux of  the J1633 optical \feii\ comparable with that of SDSS J0751+1345.   J1633 is employed in the comparison owing to its strong and narrow optical \feii\ emission lines. Also, the offset of the optical  \feii\ multiple in J1633 can be ignored (Liu et al. 2019). To verify the redshift revision, we  selected the spectral region of SDSS J0751+1345 in 4450--4700 \AA\ and 5100--5350\AA\ to cross-correlate with the spectrum of J1633. This spectral region contains the major part of  the optical \feii\ multiple and is  marked gray.  The  cross-correlation result is shown at the bottom panel of Fig.\ref{f3}.  The redshifts derived from the catalogue and the \hb\ BEL are marked with blue and red dashed lines, respectively.  Compared to the redshift of  the catalogue, the redshift revised with the \hb\ BEL is remarkably close to the peak of  the cross-correlate function  ratio (CCF) curve, which indicates that the revised redshift of \hb\ is  reliable. Thus,  in the following analysis, we consider z=1.10512 as the  systemic redshift of SDSS J0751+1345.

\section{Broad Absorption Lines}
\subsection{Absorption-free Spectrum}
\begin{figure*}
\centering
\epsscale{0.8}
\plotone{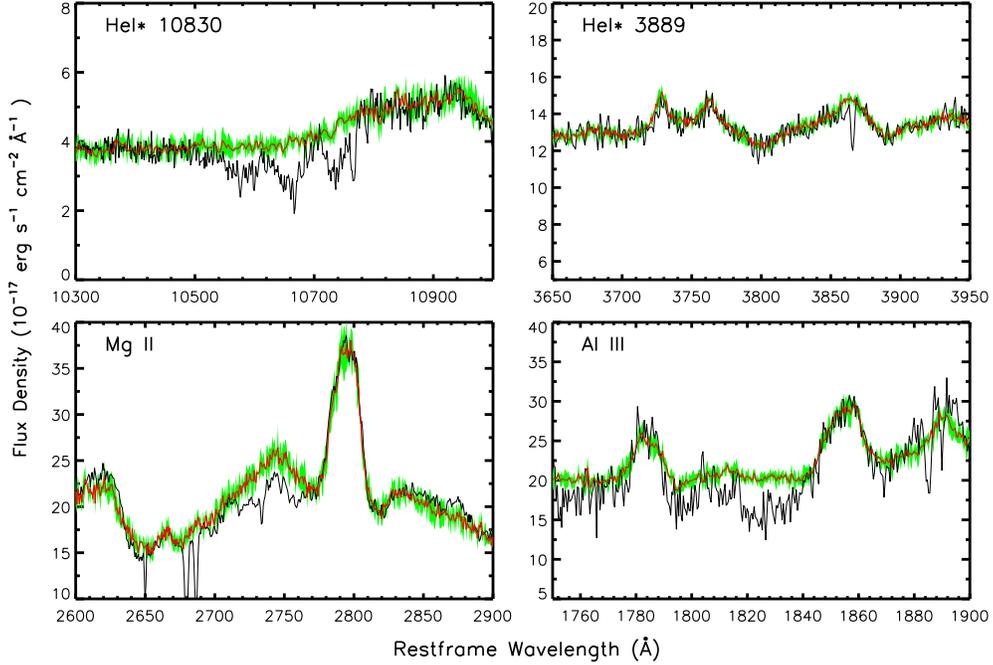}
\caption{ Pair-matching results of \heiozetz, \heiteen, \mgii\ and \aliii. In the corresponding panel, the absorption free spectrum is  shown by  a red solid line and the uncertainty is displayed by green shade.   }
\label{fmatch}
\end{figure*}

 As shown in the  inset of Fig.\ref{f1}, in the velocity space with respect to the \heiozetz\  line, we  find a prominent BAL trough located at about -10000 to -2000 \kmps, which can be identified  with  the \heiozetz\ BAL.  As another absorption line caused by the same ion, the \heiteen\ BAL   is expected to be detectable in the spectrum.  Besides, we find indications for the  \mgii\ and \aliii\ BALs. Thus,
 in this section, we will  analyze these BALs and attempt to constrain the properties of  the corresponding BAL outflow gas. For these  absorption lines in SDSS J0751+1345 , we use the pair-match method (Liu et al. 2015; Sun et al. 2017) to check  their existence and derive their absorption-free  spectra. In the case of the \heiozetz\ BAL, the spectral region of 1.02--1.10 $\mu$m is selected and the possible BAL range  of -10000 to -2000 \kmps\ with respect to \heiozetz\ is masked.   For the non-BAL quasar templates please refer to Pan et al. (2019).  These templates are  a composite of the infrared quasar spectral atlases (Glikman et al. 2006,  Riffel et al. 2006, Landt et al. 2008). The  spectral fits of three quasars in the templates are considered acceptable (reduced $\chi^2 < 1.5$, Liu et al. (2015)). The mean spectrum of the three fitted spectra can be used as the absorption-free spectrum of \heiozetz\ in SDSS J0751+1345 and the variance can be considered as the template uncertainty (Shi et al. 2016; Pan et al. 2019). The  pair-match analysis for \heiteen, \mgii\ and \aliii\ are similar  to that of \heiozetz.   For the \heiteen\ BAL, the selected spectral region is 3600 to 4000 \AA\ and  the spectral velocity range of -10000 to -2000 \kmps\  with respect to \heiteen\ is masked.  The 21 quasars in the non-BAL quasar templates are considered acceptable. For the \mgii\ and \aliii\  BALs, non-BAL templates are selected from the SDSS DR12 quasar sample (Paris et al. 2012) with spectral  S/N $ > $15 pixel$^{-1}$. These templates are required to cover the wavelength of \civ, \aliii\ and \mgii\ and  show no obvious \civ\ BAL. The selected spectral region  for \mgii\ is 2400 to 3200 \AA. Besides the spectral range of -10000 to -2000 \kmps\ in the velocity space of \mgii, the spectral wavelength range of 2675 to 2690 \AA\ in the spectrum of J0751+1345 is also masked due to possible \mgii\ absorption of  a foreground galaxy.  Seven quasars in the templates are chosen to obtain the absorption-free spectrum of \mgii.  In the case of \aliii, the selected spectral region is 1700 to 1950 \AA\ and the  corresponding  spectral velocity range of  -10000 to -2000 \kmps\ is masked.  Twelve quasars in the templates meet the criteria to be selected to derive the absorption-free spectrum. The pair-matching results of \heiozetz, \heiteen, \mgii\ and \aliii\ are shown in Fig.\ref{fmatch}.  Based on these absorption-free spectra, the corresponding EWs of \heiozetz, \heiteen, \mgii\ and \aliii\ are 55.4$\pm$12.1, 0.83$\pm$0.15, 10.8$\pm$4.8 and 6.0 $\pm$ 2.2 \AA,  respectively.  The uncertainties are at 1$\sigma$ level and include the random fluctuation in flux and the pair-matching template uncertainty. For  \heiozetz\ and \heiteen, their EWs are beyond the 5$\sigma$  significance and the existence of their BALs can be considered reliable. For  \mgii\ and \aliii, their EWs  are between 2 and 3 $\sigma$ and we can consider the existence of their BALs  to be only tentative.

(1) \heiozetz\ regime:  As shown in Fig.\ref{f4}, the \heiozetz\ BAL is blended with the HeI 10830 \& \pag\ BELs and  the velocity range of the BAL trough is from -9500 to -1800 \kmps.In the spectrum, the absorption features of \caiik\ and \naid, which can effectively  constrain  the starlight level, are not detected. Thus, the starlight from the host galaxy can be ignored. Also the flux  from the dust torus in the wavelength range of the absorption trough is very weak.   Therefore, the recovered absorption-free flux, which is shown  as the solid green line in  the corresponding panel,  is mostly contributed by the \heiozetz\ \& \pag\ BELs  and the fearless continuum.     A single  power law is employed to describe the local continuum and  for decomposition of these two components. The continuum and the BELs are displayed in the left column with blue dashed and light green dotted curves, respectively.

  (2)\heiteen\ regime: Besides the local continuum, the absorption-free flux also contains complex BELs (e.g. FeII 29 3872, FeII] 3884, \neiii). A power law, traced by the blue dashed line, is also used to model the local continuum. Different from  the \heiozetz\ BAL, only a narrow absorption trough is clearly observed in the spectrum.  In the similar velocity range, this narrow absorption structure can be also detected in  the \heiozetz\ BAL trough, which indicates that this narrow absorption trough is reliable.

  (3)\mgii\ and \aliii\ regime:  For both of these two BALs, the absorption free spectra are contributed by two components: the continuum originated from the accretion
disk, which can be fitted by a power law, and the \mgii\ \& \feii\ BELs or  the \aliii\ \& \feii\ BELs. The velocity ranges of these two BAL troughs are  from -9500 to -3500 \kmps. In this velocity range, the absorption of \heiozetz\ is obviously detected while the narrow \heiteen\ absorption trough is not.

\subsection{Measurements of $N _{\rm col} $ for the Absorption Lines}
  A comparison of the  above-metioned four BALs implies that the BAL outflow gas can be decomposed into two components. One is corresponding to the low velocity range (hereafter the LV component), which is about -3500 to -1800 \kmps. For this BAL gas, the $N_{\rm col}$ of \hei\ is thick enough to produce absorption  features (e.g. \heiozetz\ and \heiteen ). But the $N_{\rm col}$ of Mg$ ^{+} $ and Al$ ^{2+} $ in the outflow gas is so thin that their BAL troughs cannot be detected in the spectrum of SDSS J0751+1345.   The  other component corresponds to the high velocity range (hereafter the HV component)  from -9500 to -3500 \kmps. The BAL gas in this outflow component are thick enough for Mg$ ^{+} $ and Al$ ^{2+} $ to generate the \mgii\ and \aliii\ absorption  troughs. The \hei\ in this outflow component is sufficient to produce the  \heiozetz\ BAL trough. However,  due to the absorption strength ratio (gf$ _{ik} \lambda$ ) of \heiteen\ to \heiozetz\  being as small
as 0.043 (e.g., Leighly et al. 2011), this \hei\ is not thick enough to generate the detectable \heiteen\ BAL trough.

 According to the absorption  line theory, for a specified BAL, whose covering factor  $C_f(v)$ and true optical depth  $\tau(v)$, as a function of radial velocity, its normalized intensity can be  expressed as
  \begin{equation}
 I(v) = [1-C_f(v)]+C_f(v)e^{-\tau(v)}.
 \end{equation}
 $\tau(v)$ is proportional to  f$\lambda N_{\rm col}$ , where  f and $\lambda$  are the oscillator strength and rest wavelength of the BAL, respectively, and the $N_{\rm col}$ is the column density of  the corresponding absorption ion.  From this equation,  one can derive the $C_f$ and $N_{\rm col}$ of the BAL gas  if two BALs  originating from the same corresponding ion  are available, at least.

  For the LV component,  as shown in the normalized spectrum in Fig. 4, there is still  a residual flux of  around \emph{80}\%\ even at the deepest part of the \heiteen\ BAL trough . As mentioned above, the absorption strength ratio of \heiozetz\ to \heiteen\ is about 23.3. Also, in the \heiozetz\ trough, we detected  significant residuals after removal of \heiozetz\ \& \pag\ BELs.  These facts imply that this  BAL component only partially covers the accretion disk.
We  use the \heiteen\ and \heiozetz\ absorption lines that  transition from the same energy level of \hei\ to derive $C_f$ and $N_{\rm col}$. The $N_{\rm col}$ of \hei\  is derived as 3.5 $ \pm $ 0.5$ \times $ 10$ ^{14} $ cm$ ^{-2} $.
The \mgii\ and \aliii\ BALs in this component is undetected. With the $C_f$ of \hei\, we
 derive the 3-$ \sigma $ upper limit on the column density for the two ions,
 7 $ \times $ 10$ ^{13} $ cm$ ^{-2} $ and 8 $ \times $ 10$ ^{13} $ cm$ ^{-2} $, for Mg$ ^{+} $ and Al$ ^{2+} $, respectively.

\begin{figure*}[htbp]
\centering
\epsscale{1}
\plotone{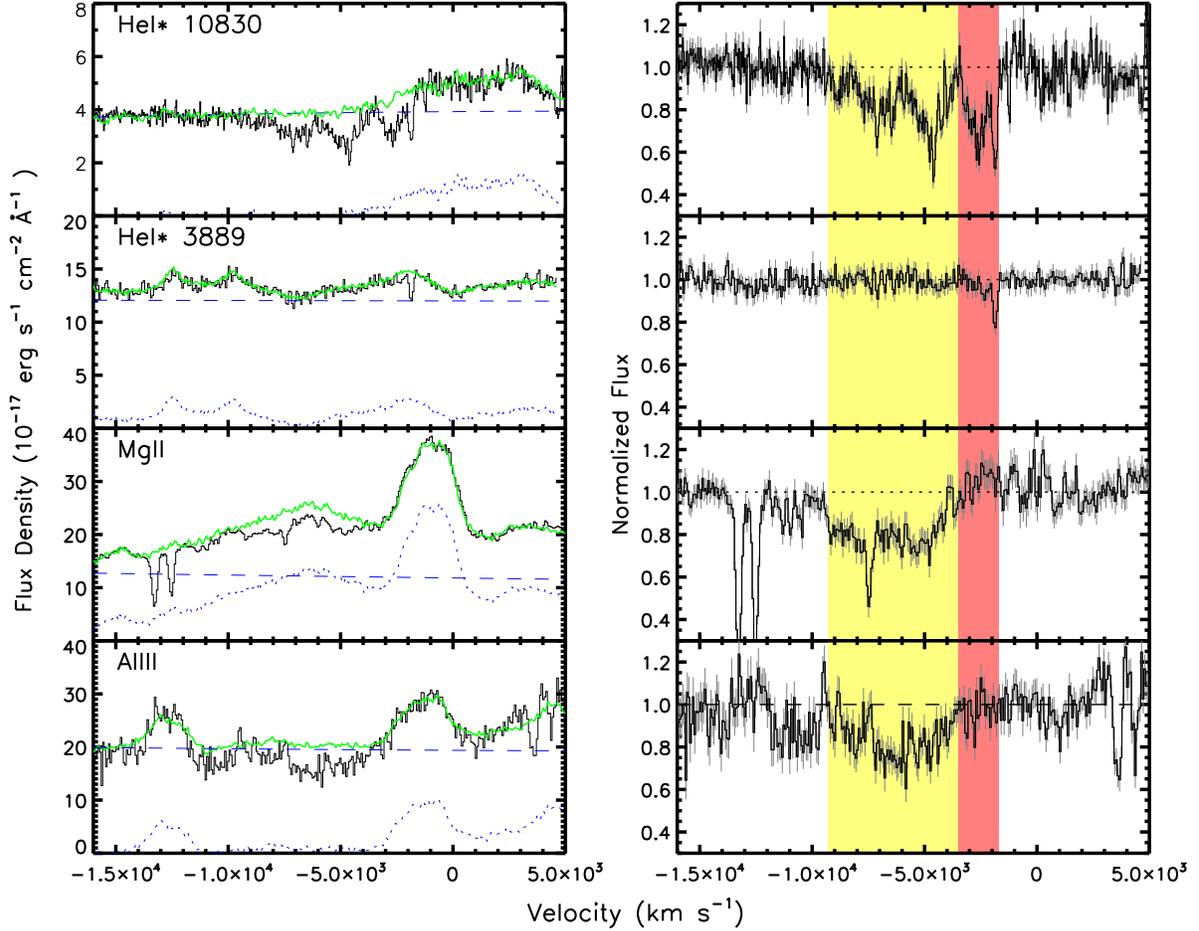}
\caption{\textit{Left}:  Pair matching results and normalized spectra of \aliii, \mgii, \heiteen\ and \heiozetz.
The pair matching results are decomposed into the continuum (blue dashed) and the BEL (green dotted).
Based on the assumption that the accretion disk is partially obscured by the BAL gas,
the continuum is considered as the absorption-free spectrum.
\textit{Right}: The BALs after the subtracting of the absorption-free spectrum in the velocity space.
The BALs can be divided to two components: High-velocity component (HV, yellow region) and Low-velocity
component (LV, red region). The \aliii\ and \mgii\ BALs are mainly contributed by the HV component
and the \heiteen\ is mainly contributed by the LV component. The \heiozetz\ BAL contains both HV and LV component.
   }
\label{f4}
\end{figure*}

  For the HV component, according to  Fig.\ref{f4}, the residuals of \heiozetz, \mgii\ and \aliii\ BALs are obvious after   removal of the  BELs in their BAL velocity ranges.  This indicates that
  these ions in the BAL gas  are optical thin or the BAL gas partially obscures the accretion disk.
  We first assume that the BAL gas partially  obscures the  accretion disk.
  Thus, the BELs in the velocity ranges of the BALs should be removed from the absorption
  free spectra and
  the absorption profiles of \heiozetz, \mgii\ and \aliii\ are displayed in Fig.\ref{f4} (right).
  In order to measure the $N_{\rm col}$ of these ions in  the HV component,  we first assume the $C_f$ is constant in the velocity range of  the HV component. As the two lines  are produced by the transitions from the same energy level,
   \heiozetz\ and \heiteen\  are employed to constrain  $C_f$. Different from  the case of the LV component,
  the \heiteen\ BAL in the HV component is undetected. Thus,  these two lines can be used to provide only a lower limit of $C_f$. As mentioned above,  the absorption strength ratio (gf$ _{ik} \lambda$ ) of  \heiozetz\ to \heiteen\ is as large as 23.3. For the unsaturated pixels in the \heiozetz\ HV BAL velocity
  range, the absorption at  the corresponding pixels of \heiteen\ is undetectable,  as can be seen in the observed
  spectrum. Therefore, to obtain the lower limit of $C_f$, we can decrease  $C_f$ from 1
  to  where  some pixels start to be saturated in \heiozetz\ HV BAL velocity range.

\begin{figure*}[htbp]
\centering
\epsscale{1}
\plotone{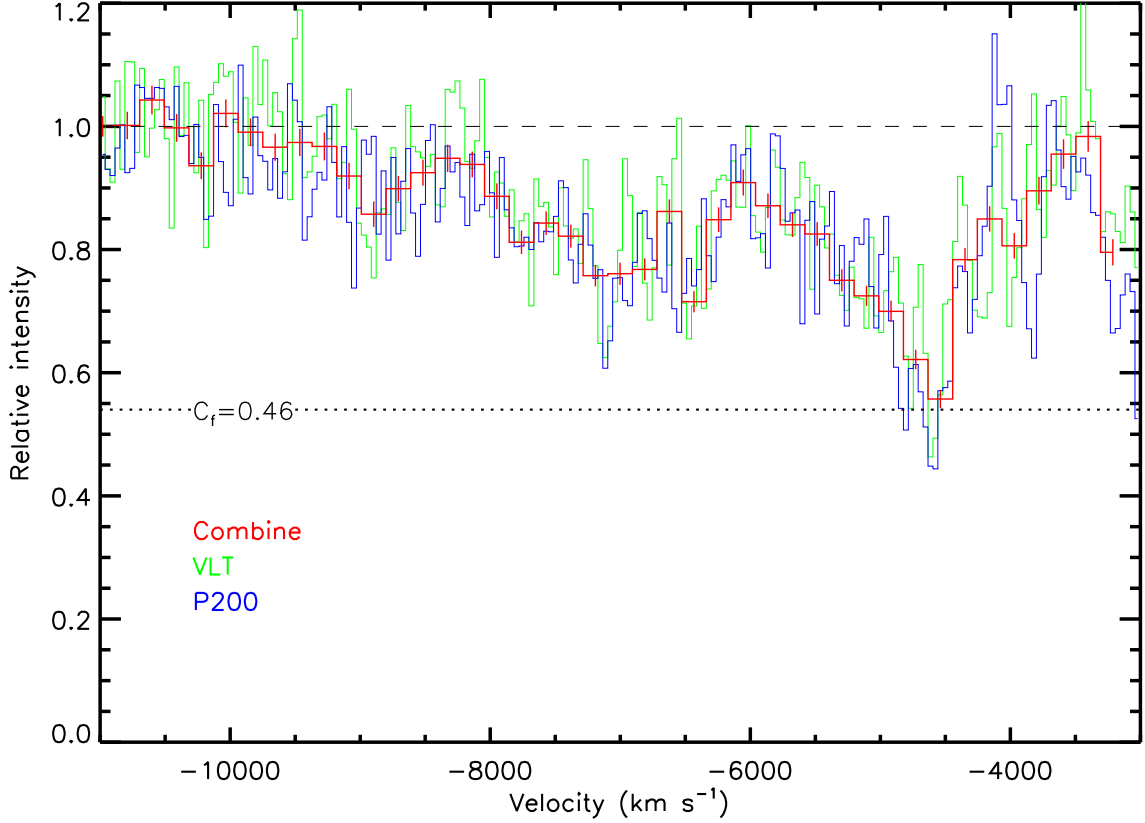}
\caption{ Comparison of the \heiozetz\ BAL profiles between the P200 spectrum (blue) and the VLT spectrum (green).
	 This indicates that the deepest absorption structure near -4500 \kmps\ is reliable and
	the constant $C_f$ of  the HV component can be derived from this absorption structure.
	To further  conservative estimate  the $C_f$, we  rebin the composite spectrum to the
	resolution of R =1000 , which is displayed in red. The lower limit of $C_f$ can be obtained from the
	deepest pixel near -4500 \kmps\ which is $\sim$0.46 and marked by the horizontal dotted line.
   }
\label{f5}
\end{figure*}

\begin{figure*}[htbp]
\centering
\epsscale{1}
\plotone{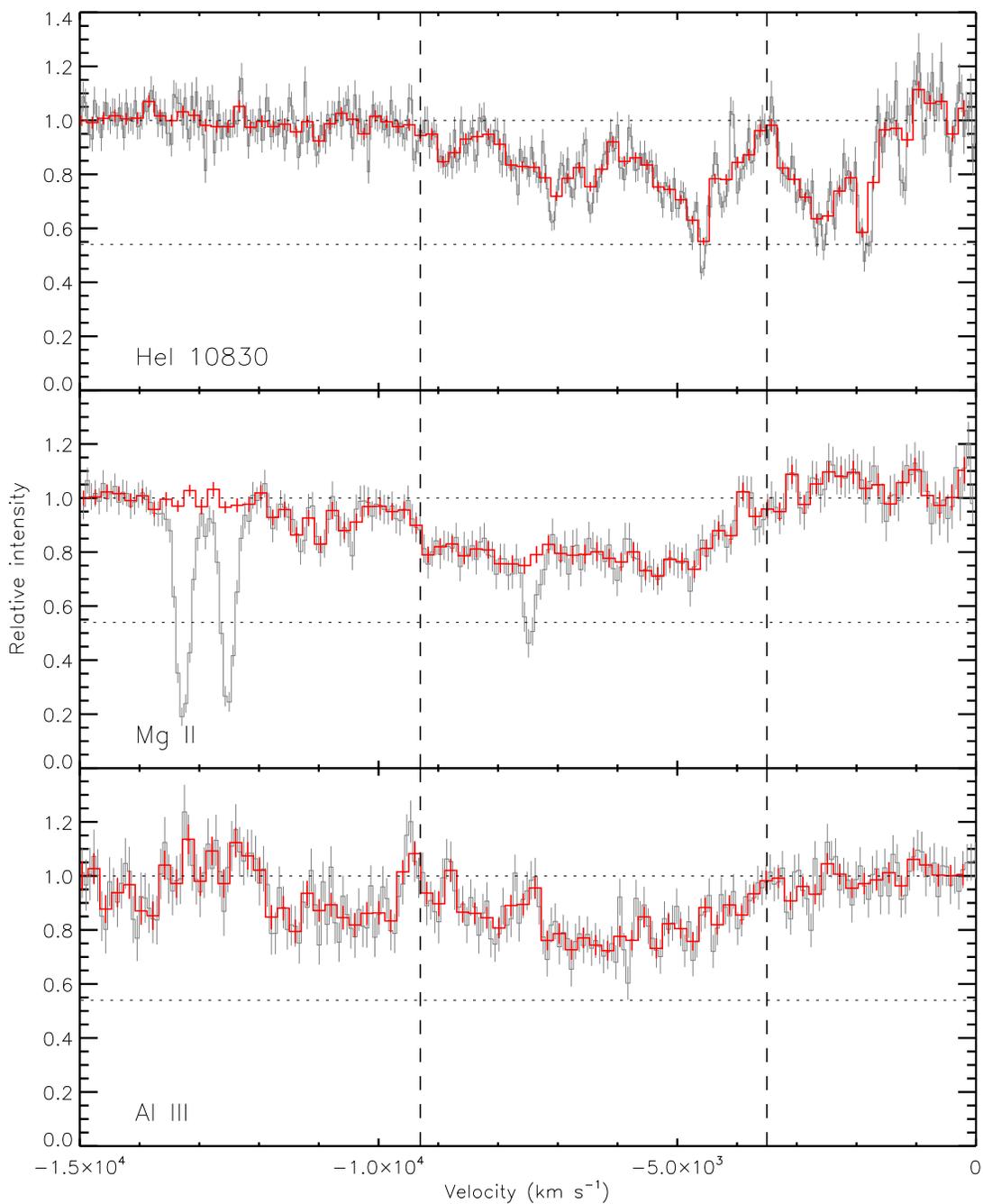}
\caption{ Estimation of  the column density on the \hei, Mg$ ^{+} $ and Al$ ^{2+} $ through
	the \heiozetz, \mgii\ and \aliii\ HV BAL. These BAL profiles are all rebinned to the
	resolution of R=1000 and displayed in red.  $C_f$ is assumed as constant in the
	velocity range of  the HV component. The lower limit of $C_f$, 0.46, is marked with
dotted lines and its upper limit is chosen as 1.    }
\label{f6}
\end{figure*}

  It should be noted that,  through this method, the lower limit of $C_f$ is determined by the
  deepest pixels in the \heiozetz\ HV component and these pixels may be affected by the
  imperfect spectral observations and reductions.
  We used two methods to examine the reliability of  the measurements.
  First, there are two spectroscopic observations (P200 and VLT spectra)  in the NIR band
  where the \heiozetz\ HV BAL  is located.  By the comparison of the two spectra,
  the effect of observations and reductions at the deepest pixels can be investigated.
  As shown in Fig.\ref{f5}, the \heiozetz\ BALs extracted from the VLT spectrum and P200 spectrum are
  shown in green and blue, respectively. It can been seen that the deepest  the pixels of  the BAL  are located   at about -4500 \kmps\ and the absorption structures are nearly the same.
  This indicates that this absorption structure is reliable.
  Second, to further investigate the observational effect and  to estimate   more conservatively the lower limit of $ C_f $, we  produce a
  composite spectrum and  rebin it to a lower resolution of R = 1000  as shown
  in Fig.\ref{f5} (red curve). This composite spectrum is employed to obtain the
  lower limit of $C_f$. We  find that when  $C_f$ decreases to 0.46, the pixels
  of  the \heiozetz\ BAL at about -4500 \kmps\ start  to become saturated\footnote{According to Eq.1,
  when the normalized residual flux I(v) of a pixel is equal to $C_f$,
  the absorption of this pixel is saturated and  $\tau$ is not measurable.
  We can use  Eq.1 to estimate  $\tau$ of  the pixels in the BAL, and identify
  the criterion of saturation as I(v)+$ \sigma $(v) = $C_f$, where  $ \sigma $(v)   is the error of I(v) at the corresponding velocity. }  and hence $C_f$ = 0.46 can be considered
  as its lower limit. We also  rebin the \mgii\ and \aliii\ BALs to R=1000  as shown
  in Fig.\ref{f6}. Compared to the lower limit of $C_f$, both the \mgii\ and \aliii\ BALs
  are unsaturated. Thus, we can obtain the $N_{\rm col}$ upper limits of \hei,
  Mg$ ^{+} $ and Al$ ^{2+} $.  Besides, the lower limits of $N_{\rm col}$  on \hei, Mg$ ^{+} $
  and Al$ ^{2+} $ can be derived assuming $C_f$=1. Based on the range of $C_f$,
  we can derive the $ N_{col} $ ranges of \hei, Mg$ ^{+} $ and Al$ ^{2+} $ ,  to be 0.8--1.6
  $ \times $ 10$ ^{14} $,  2.1--3.6 $ \times $ 10$ ^{14} $  and 2.8--5.5 $ \times $ 10$ ^{14} $ cm$ ^{-2} $, respectively.
   Note that in the above analysis, we assumed that the different BALs share the same covering factor to the accretion disk.   However, according to the  disk model, the size of  the disk, where the photons  of the   corresponding wavelength of \heiozetz\  is emitted, is about 4 times larger than that  of the  photons of the corresponding  wavelengths of \aliii\ and \mgii.
  Therefore, the covering factors of  the \aliii\ and \mgii\ BALs should be larger than that of  the \heiozetz\ BAL. Thus, our assumption of the lower limits of \aliii\ and \mgii\ covering factors  of 0.46 can be considered conservative and the calculation based on this assumption is  available.

  \section{Ionization Model for The BALs}
  For the BAL outflow of SDSS J0751+1345, we can make a natural assumption that there are two absorbers, which is consistent with the different velocity ranges of HV and LV components.
   One of the absorbers is corresponding the LV BAL components. This absorber has slower velocity, smaller column density of  Mg$ ^{+}  $ and Al$ ^{2+} $ but larger column density of \hei\ than the other. Another absorber is faster, with thicker Mg$ ^{+}  $ and Al$ ^{2+} $ but thinner \hei\ than the former.
  Based on $N_{\rm col}$ of different ions, we attempt to constrain the characteristics of these
  two absorbers with the help of photoionization model.

   We employ the soft package CLOUDY  (c13.03; Ferland et al. 1998) to  construct the photoionization model of the BALs in SDSS J0751+1345 .
   In photoionization simulations,  the absorbed gas  is assumed to be slab-shape, dust-free and solar elemental abundance.  For simplicity  of computation, the density, metallicity and abundance in the gas-slab  are assumed to be uniform.  It should be noted that \hei,  Mg$ ^{+}  $ and Al$ ^{2+} $ have different ionization potentials. The ionization potential ranges of \hei,  Al$ ^{2+} $ and Mg$ ^{+}  $ are 24.4 to 54.4 eV, 18.8 to 28.4 eV and 7.6 to 15.0 eV, respectively. The ionization potential range of Mg$^+$ has no overlap with  those of \hei\ and  Al$ ^{2+} $. Also, with a gas slab, the main ionization zone of Mg$^+$ is beyond the ionization front where the \hei\ and Al$^{2+}$ are ionized. Nonetheless, for simplicity, we assume the same density, metallicity and abundance for the absorbers of   \hei,  Al$ ^{2+} $ and Mg$ ^{+}  $.
   This medium is ionized by continuum  originating from the the central engine, whose SED is in the form that is defined  by  Mathews \& Ferland (1987, hereafter MF87).
   The simulation results are  compared with the $N_{\rm col}$ of the
   ions in LV or HV BAL components.

\begin{figure*}[htbp]
\centering
\epsscale{1}
\plotone{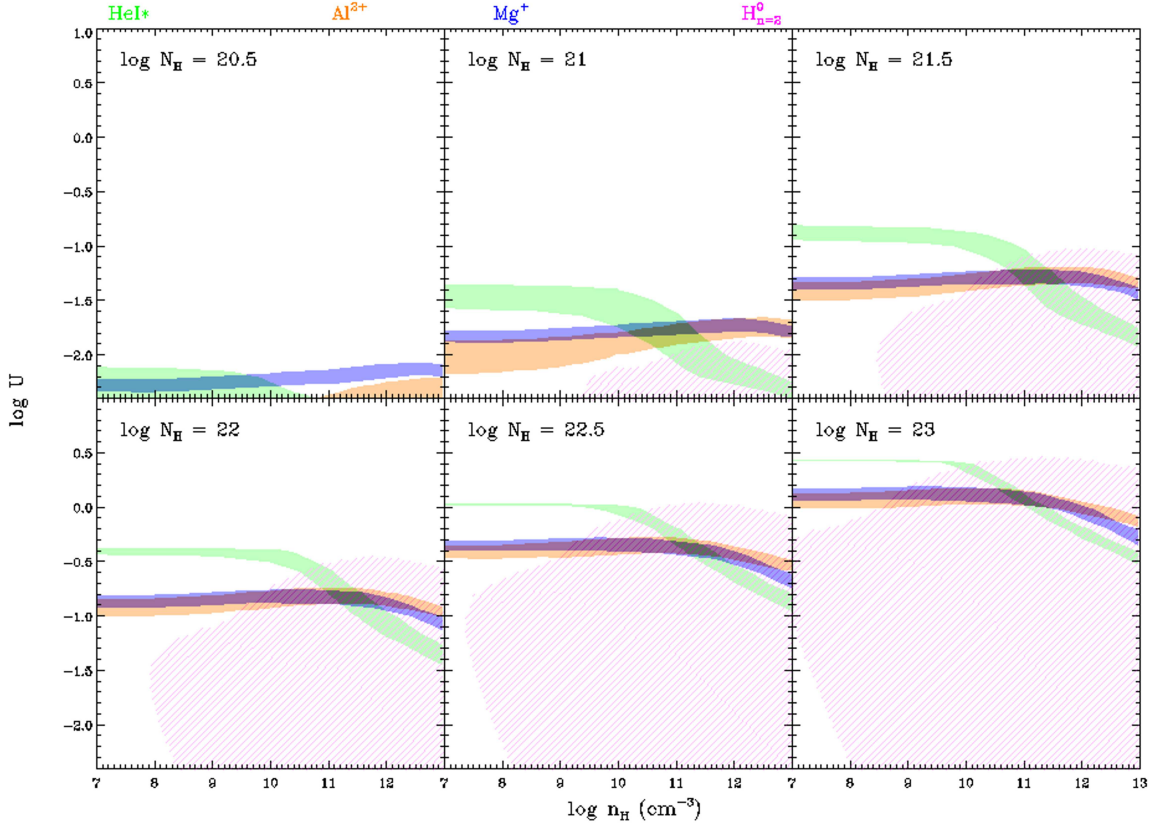}
\caption{ Allowed parameter intervals of Mg$ ^{+} $ (blue), Al$ ^{2+} $ (orange)  and \hei\ (green)
	of HV BALs  on the plane of density $n_{\rm H}$ versus ionization parameter $U$ as calculated by CLOUDY simulations
	for the column density $N_{\rm H}=10^{20.5}-10^{23}$ cm$^{-2}$, solar abundance, and MF87 SED.
	The excluded areas of H$ ^{0}_{n=2} $ 3-$\sigma$ upper limits are marked with magenta lines.
	It can be seen that, at  $N_{\rm H} = 10^{21}$ cm$^{-2}$, all the allowed parameter intervals  overlap.}
\label{f7}
\end{figure*}

   For  the HV component, with our measurement for the $N_{\rm col}$ of \hei,  and a assumption of the upper limit for the density ratio of HeI$^*$ to He$^+$ ( Arav et al. 2001; Ji et al. 2015)  the lower limit for the He$^+$ column density in the outflow can be obtained as $\sim3\times10^{19}$ cm$^{-2}$.  Assuming solar abundance, the lower limit for the H II column density, also the minimal hydrogen column density ($N_{\rm H} $) of the HV absorber, can be estimated as $\sim3\times10^{20}$ cm$^{-2}$. Thus, in the simulations,
 we set an array of $N_{\rm H} $ from $10^{20.5}$ to $10^{23}$ cm$^{-2}$  with a step size of 0.5 dex.
 For each value of $N_{\rm H} $, we run over the grids of simulations with different ionization parameter and hydrogen density. The range of ionization
parameter  is -2.4$ \leq $ log $U$ $ \leq $ 1 and the  hydrogen density is 8 $\leq $ log $n_{\rm H} $ $ \leq $ 12. The steps for both of these two parameters are 0.2 dex.
The allowed $n_{\rm H} $ and $U$ intervals   on the $n_{\rm H} $-U plane at specific  $N_{\rm H} $ can be derived through the comparison between the column densities of \hei, Mg$ ^{+} $, and Al$ ^{2+} $ and those simulated by CLOUDY.
As shown in Fig.\ref{f7}, for the constant log $N_{\rm H} $, the corresponding area in the $n_{\rm H} $-U plane of \hei, Mg$ ^{+} $, and Al$ ^{2+} $ are marked  green, blue, and orange, respectively.
Besides the measured \hei, Mg$ ^{+} $, and Al$ ^{2+} $ column densities in the HV absorber,
the non-detection of Balmer BALs can be used to constrain the allowed parameter intervals.
With the lower limit of  the covering factor to the accretion disk $ C_f $ = 0.46, we can derive
the 3-$ \sigma $ upper limit of H$ ^0_{n=2} $ column density as 2.8 $ \times $ 10$ ^{14} $ cm$ ^{-2} $.
The parameter area with  magenta solid line  is excluded from the 3-$ \sigma $ upper limit of H$ ^0_{n=2} $ column density in Fig.\ref{f7}.
Only in the case of log $N_{\rm H} $ (cm$ ^{-2} $) = 21, the four allowed parameter areas  overlap. In summary,
the simulation results suggest the physical parameters of log $n_{\rm H} $ (cm$ ^{-3} $)from 10.3 to 11.4,
the log $U$ from -1.83 to -1.72, and log $N_{\rm H} $ (cm$ ^{-2} $) $\sim$ 21 for the HV outflow gas.

\begin{figure*}[htbp]
\centering
\epsscale{1}
\plotone{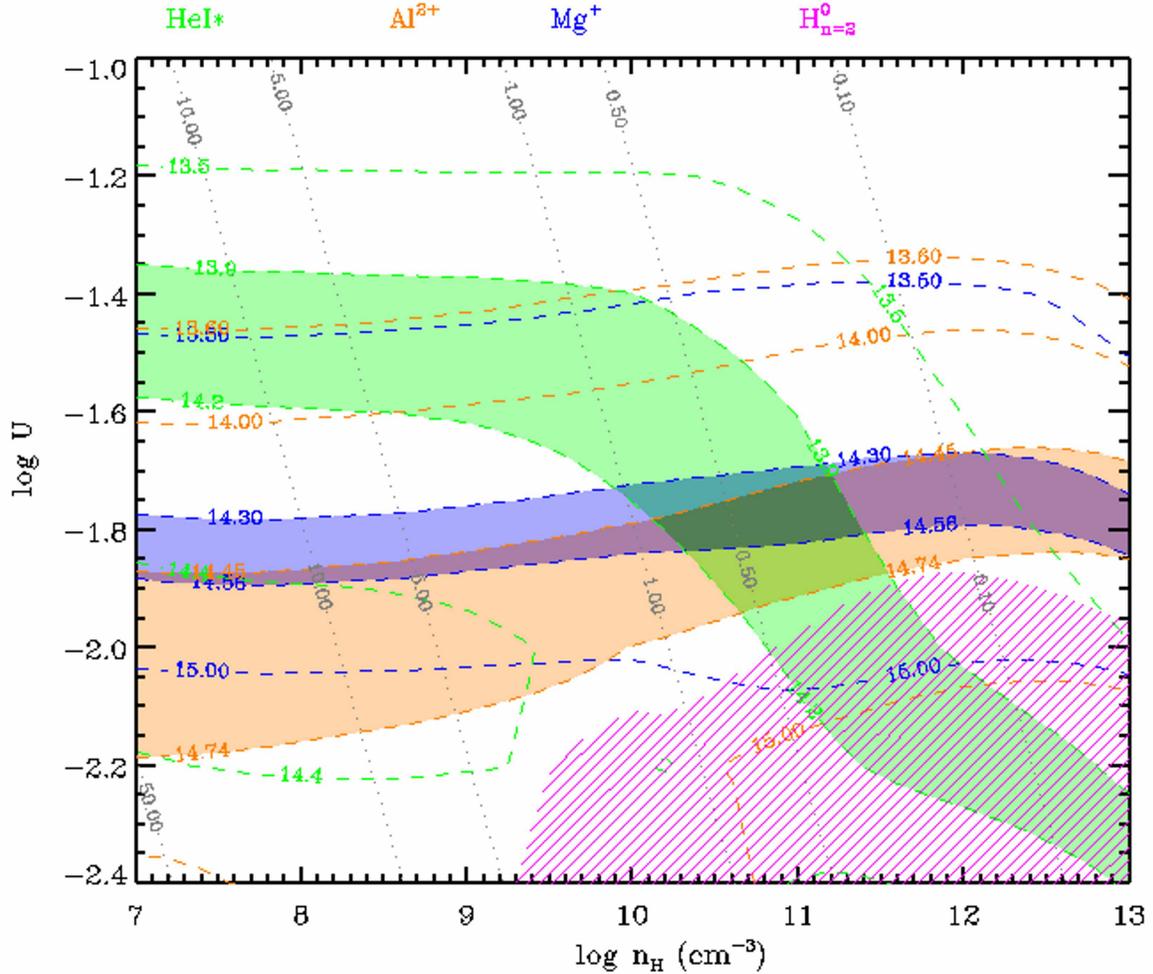}
\caption{ As  in Fig.\ref{f7}, the detailed contours of Mg$ ^{+} $ (blue), Al$ ^{2+} $ (orange)
	and \hei\ (green) of  the HV BALs  on the plane of $n_{\rm H}-U$.  The areas marked by magenta lines are excluded by the upper limit of H$ ^{0}_{n=2} $ column density
The distance to the central engine is also contoured with dotted lines. The overlap region
corresponds to 0.5 pc, which is consistent with the size of  the BLR.    }
\label{f8}
\end{figure*}

  With the constrained parameter intervals, we can place a limit of the distance between  the HV absorber
  and  the central engine.  According to the definition of U, we can  express the distance between  the absorbing medium and  the central engine as  $R_{abs} ~=~ (Q(H)/(4 \pi cUn_e) )^{0.5}$, where Q(H), the number of ionizing photons, can be derived as $Q(H) = \int_{\nu}^{\infty}{L_{\nu}/h \nu d \nu }$. For our source, the monochromatic luminosity at 5100\AA ($\lambda L_{\lambda} (5100\AA)$) can be obtained through the spectrum as $\sim$$7.0 \times 10^{45} ~ erg~ s^{-1}$.  Based on the MF 87 SED,
  we derived $\rm Q(H) \approx 5.9 \times 10^{56}~ photon~ s^{-1} $.
  Thus, $R_{abs}$ can be calculated, which is  represented by gray dashed lines  on the $n_{\rm H} $-U plane
  (Fig.\ref{f8})
  Compared with the contoured lines of $R_{abs}$, the  location of HV BAL absorber is about 0.5pc away from the central engine.
   Furthermore, through the $\lambda L_{\lambda} (5100\AA)$ of our source, we can also estimate the radius of BLR, R$ _{BLR} $, according to the equation
 \begin{equation}
 R_{BLR} = 32.9_{-1.9}^{+2.0}[\frac{\lambda L_{\lambda} (5100\AA)}{10^{44} ~ erg~ s^{-1}}] ~~ lightdays,
 \end{equation}
 which is fitted with the  reverberation mapping results in Kaspi et al. (2000). Thus, for our source, R$ _{BLR} $  is calculated as about 0.5pc, which is very similar to the distance of  the HV BAL absorber.

Assuming  the solar abundances and the maximum \hei\/He$^+$ abundance ratio, we obtained an minimal  column density of the LV absorber, $N_{\rm H}$ $>$ 10$^{21}$ cm$^{-2}$,  which is larger than that of the HV absorber. For a  gas slab with increasing thickness illuminated by a quasar ionization continuum radiation, we expect to detect absorption in the sequence of \heiozetz, \aliii\ and \mgii, which have similar oscillator  strengths and decreasing ionization potentials (e.g. Ji et al. 2015; Liu et al. 2016). For  the LV component, the  HeI$^{*}$ $\lambda$3889,10830 BALs are significantly detected while neither \aliii\ nor \mgii\ is detectable.  This indicates that the column desity of LV BAL gas is large enough for the development of  the He$^+$ zone, and yet not enough for the Al$^{2+}$ zone (let  alone Mg$^+$ zone), and therefore has a higher ionization level than the HV absorber with a significant detection of  the \aliii\ and \mgii\ lines. This, incorporating the fact that the LV absober gas has a higher column density than the HV absorber, indicates that either the  LV absorber has  a much lower density than  the HV absorber, or the LV absorber is located much closer than the HV absorber to the central engine, since both of the LV and HV absorbers are illuminated by the same continuum source. It is  hard to  imagine, if not impossible, that, for two absorption  gas clouds in the nuclear region of the same quasar, a higher column density (LV) absorber has a lower density while a lower column density (HV) absorber has a higher density.  Hence the low density scenario for  the LV absorber is highly disfavored. Conservatively assuming that both LV and HV absorbers have the same density , we  derive a maximal  distance of the LV absorber $R_{LV}$ $<$ 0.5 pc. The  actual value should be much smaller than this, since both  the density and column density of the LV absorber may be larger than that of the HV absorber. In this scenario, the LV absorber could be naturally taken as the headstream of the HV absorber (Hall et al. 2011)  and the outflow is accelerated.  Further high quality spectroscopy is needed to confirm such a  speculation, by  imposing more stringent constraints on  \aliii\ and \mgii\ BALs and  detecting of more absorption lines from other ions.

\section{Blueshifted Component in BELs}
Besides the BALs, blueshifted  BELs are also a significant feature of the outflow. For SDSS J0751+1345, the main relatively independent BELs expected are \aliii, \mgii, \hb, \hei\ 10830 and \pag.
The blue edge of \hei\ 10830 and \pag\ is polluted by the LV BAL of \heiozetz\ and
the existence of  any blueshifted BEL component is difficult to be determined.
The model-fitting to the \hb\ in Section 2 revealed no evidence of the blueshifted component.
In this section, we  mainly focus on the \aliii\ and \mgii\ to determine whether the
blueshifted BEL components  are present.

\begin{figure}[ht]
\epsscale{1.1}
\plotone{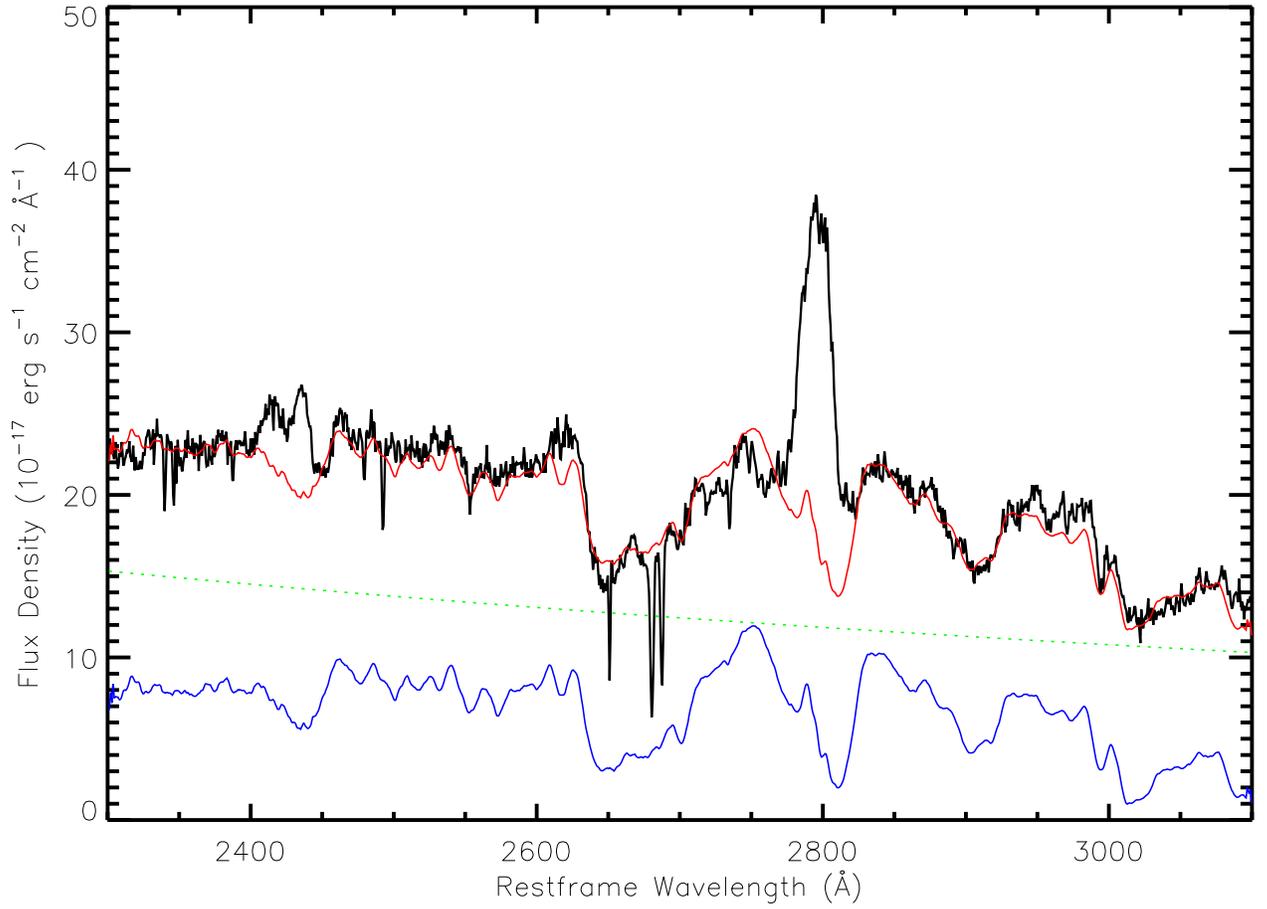}
\caption{ Fitting result of UV \feii\ multiples. The continuum is plotted by the green dotted line.
	The  \feii\ components are displayed by  the blue curve and  the sum in red.
}
\label{f10}
\end{figure}

For \aliii, due to the effect of  the \aliii\ BAL, the local continuum is difficult to  determine directly from the spectrum. However, in Section 4.2, we have obtained the local continuum of  the \aliii\ absorption free spectrum. Thus, we can use this continuum as the local continuum of \aliii\ in SDSS J0751+1345 and subtract this continuum from the observation spectrum to obtain the \aliii\ BEL.
To derive the \mgii\ BEL,  a power law and a Gaussian-kerneled  UV \feii\ template  (Tsuzuki et al 2006) are employed to  fit the continuum and the UV \feii\ multiples around \mgii\ in SDSS J0751+1345.
 The \feii\ template (Tsuzuki et al 2006) broadened with a Gaussian-kernel  is used to fit the
\feii\ multiples around \mgii\ in SDSS J0751+1345. The fitting results are displayed in
Fig.\ref{f10}.
 As suggested by the Wang et al. (2011), the parameter BAI is a good indicator for the existence of  a blueshifted component in the BEL
A value of BAI$>$0.5 would suggest the existence of  a blueshifted BEL component.
For  the \mgii\ doublet,  in light of the line core of IZW1, we set its rest-frame wavelength  to 2999.4 \AA. The rest-frame wavelength of \aliii\ doublet is chosen
as 1859.4 \AA, corresponding to the line ratio of \aliii\ 1857 / \aliii\ 1863=1.25.
The BAI of \mgii\ and \aliii\ are calculated  to be $0.60 \pm 0.01$ and $0.67 \pm 0.01$, respectively.
Note that  if possible existence of NELs in the two lines  is considered, these values of
BAI  would be even be larger. This indicates that the blueshifted components are detected in the
\mgii\ and \aliii\ BELs.

\begin{figure*}[htbp]
\centering
\epsscale{0.8}
\plotone{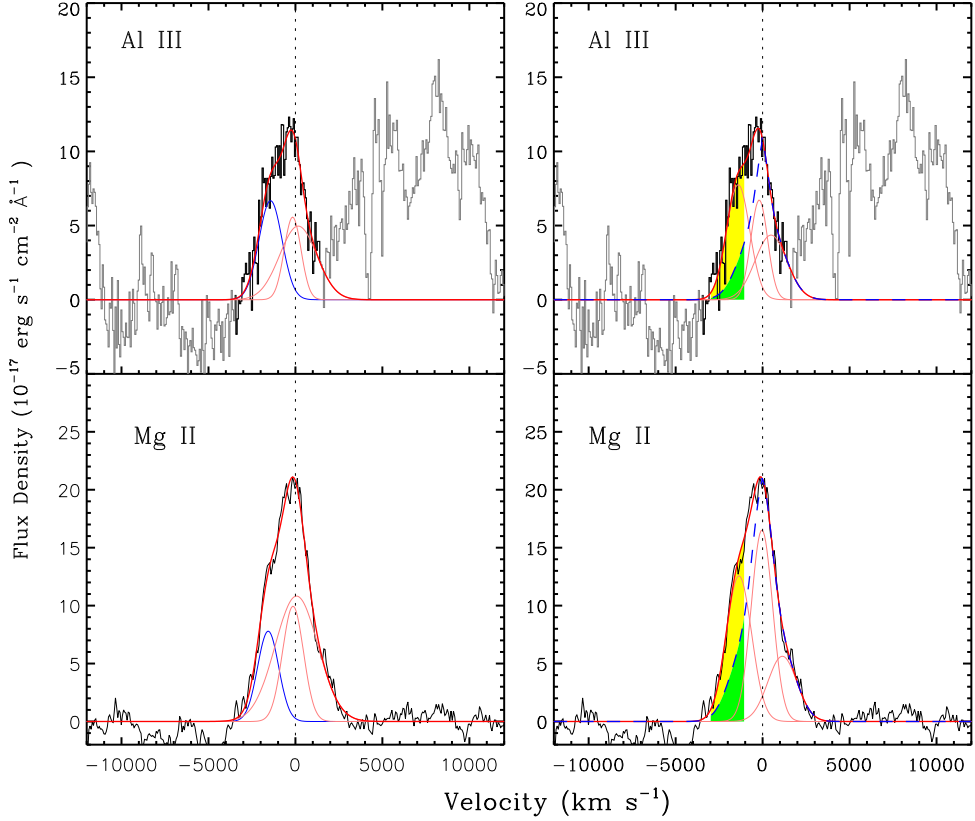}
\caption{\textit{left}: Decompositions of \aliii\ and \mgii. The BLR components are fitted with two Gaussians and displayed by pink curves.  The profile of the blueshift components in the two lines are  tied and described by one Gaussian and displayed by blue curves.  The line ratio \aliii/\mgii\ for the blueshifted component is obtained as 0.56$\pm$0.02. \textit{right}: Estimation of  the lower and upper limits of \aliii\ and \mgii\  BELs from the outflow.
	For each of \aliii\ and \mgii, the flux in the wavelength range between -3000 and -1000 \kmps\
	comprises the emission from  the BLR  and hence
	 sets an upper limit of the intensity of the blueshifted component.
	Similarly, the mirror symmetric flux from -3000 to -1000 \kmps\ (green region)
	contains the redshifted component of  the BEL,  and hence the subtraction of total flux (yellow
	region) gives lower limit of the intensity of the blueshifted component.   }
\label{f11}
\end{figure*}

We first  try to decompose the \aliii\ and \mgii\ BELs into two components: one is blueshifted and emitted
    from the outflow,  and the other is in the quasar's rest-frame from the normal BLR. The blueshifted component  is modelled with one Gaussian,
    while the component from  the normal BLR  is fitted with two Gaussians. The profiles and intensities of the BLR Gaussians  are free  except that their peak offsets are limited in the range of -150 to 150 \kmps, due to the uncertainties of wavelength calibration and redshift determination. The blueshifted Gaussians in \aliii\ and \mgii\ share the same profile. The fitting results are shown in the left panel of Fig.\ref{f11}.  The blueshifted  line ratio of \aliii/\mgii\ can be derived as 0.56$\pm$0.02. However, the blueshifted component and the BLR component in specific emission line are heavily blended, especially in \mgii,  impling that these decompositions are model-dependent.

Thus, we  attempt to constrain the corresponding upper and lower limits of  the blueshifted line ratio of \aliii/\mgii\ without  line decomposition.
For each of \aliii\ and \mgii, we first use three Gaussians to describe its emission profile
and plot the results in Fig.\ref{f11}  as red solid curves.
The velocity range of -3000 to -1000 \kmps\ is chosen to calculate the blueshifted line
emission ratio of \aliii/\mgii.
The lower limit of -3000 \kmps\ is set because it is near the blue ends of
\mgii\ and \aliii\ BELs and the upper limit of -1000 \kmps\ is set to minimize the
contamination from the narrow component of  the emission line.
In this velocity range, the emission of \aliii\ and \mgii\ contains the blueshifted component
and the normal BEL,  and hence  their total  intensities can be considered as the upper  limits of  the
\aliii\ and \mgii\ blueshifted  components, respectively.
Under the assumption that the normal BELs in the broad emission lines arise from the BLR
for which the predominant motion is either Keplerian or virial (see Gaskell 2009 for a review),
the \aliii\ and \mgii\ normal BELs are expected  to be symmetric to the line centroid at the
rest-frame wavelength.
The lower limit  on blueshifted \aliii\ and \mgii\ BEL can be estimated by subtracting the
blue-side symmetric flux from the total (yellow shaded region in the right panel  of Fig.\ref{f11} ).
Thus, to be  conservative, we  estimate the upper limit of  the line ratio of
\aliii\ to \mgii\ through the upper limit of  the \aliii\ blueshifted component divided by the lower limit of  the \mgii\ blueshifted
component. Similarly, 
the lower limit of  the \aliii\ blueshifted component divided by the upper limit of  the
\mgii\ blueshifted component gives the lower limit of the \aliii/\mgii\ ratio.
The lower and upper limit  of the line ratio of  the blueshifted \aliii\ to \mgii\ BEL
is calculated  to be 0.25 and 0.93, respectively.

\begin{figure*}[htbp]
\centering
\epsscale{1}
\plotone{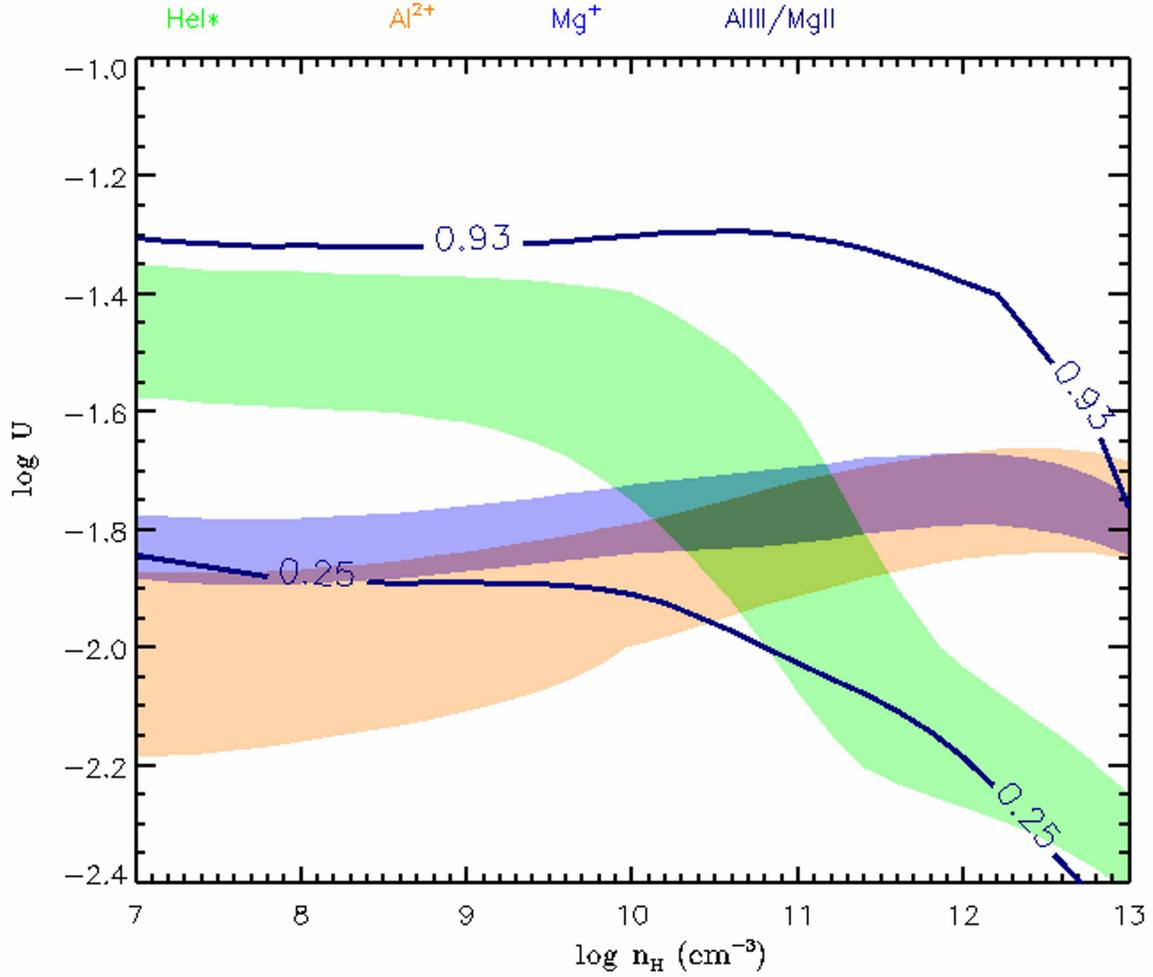}
\caption{ Like Fig.\ref{f8}, the $n_{\rm H}-U$ contours for the corresponding line ratio of \aliii/\mgii\ (navy line),
	which cover the allowed $n_{\rm H}-U$ parameter ranges derived from the HV BAL.  It implies
that the outflow  which the blueshifted \aliii\ and \mgii\ BELs  originate from
and that  produces the HV BALs are not isolated.     }
\label{f12}
\end{figure*}

Although the BEL outflow is only detected in \aliii\ and \mgii\ in SDSS J0751+1345,
we  investigate whether the parameters constrained by the BALs can reconcile with the line ratio of
\aliii/\mgii\ via CLOUDY simulations. As the parameters of LV BALs cannot be fully determined,
we  attempt to reproduce the \aliii/\mgii\ ratio with the parameters of HV BALs.
The simulation settings are the same  as those of HV BALs except  for $N_{\rm H}$ that is fixed to be
10$ ^{21} $ cm$ ^{-2} $ (corresponding to the derived $N_{\rm H}$ for HV BALs).
The parameter intervals  on the $n_{\rm H}$-$U$ plane allowed by the \aliii/\mgii\ ratio are
delineated by  the navy solid lines in Fig.\ref{f12} and they include the overlapped area
derived from HV BALs.
This implies that the HV BAL outflow and the blueshifted BEL outflow may not be independent.
In fact, with $N_{\rm H}$ = 10$ ^{21} $ cm$ ^{-2} $, log $U$ = -1.77 and log $ n_{\rm H} $ (cm$ ^{-3} $) = 10.8,
the best physical parameters inferred for the HV BAL, we can derive the EWs of \mgii\ and \aliii\
from CLOUDY simulations, which  are about 15\AA\ and 4\AA\, respectively.
According to the analysis above, the lower  limits of the observed
 EWs of the \mgii\ and \aliii\ blueshifted component can be estimated as 8 and 2\AA, respectively,
which  are consistent with the values  of the simulations,
suggesting again that the BAL and BEL  outflows may be associated.

\section{Summary and Discussion}
We make a detailed analysis  of the characters of BAL and BEL outflows of SDSS J0751+1345.
With the broad \hb\ and \oii\ NEL, we revise the systematic redshift  to be z=1.10512.
The \aliii, \mgii, \heiteen\ and \heiozetz\ BALs are detected in the spectrum,
suggestive of  AGN BAL outflows. The analysis to the velocity ranges of the BALs indicates that
there are two BAL components co-exiting in SDSS J0751+1345, namely  a HV and  a LV  component.
 The HV BAL component is detected in \heiozetz, \mgii\ and \aliii\ and
its velocity range is from -9300 to 3500 \kmps. The covering factor of  the HV component can be constrained
in the range  of 0.46 to 1,
and the $N_{\rm col}$ of \hei, Mg$ ^{+} $ and Al$ ^{2+} $ can be derived as
0.8--1.6 $ \times $ 10$ ^{14} $,  2.1--3.6 $ \times $ 10$ ^{14} $  and 2.8--5.5 $ \times $ 10$ ^{14} $ cm$ ^{-2} $, respectively.
By  comparing the observations with the photoionization simulations,
 $n_{\rm H}$ and $U$ for the HV BAL component  are constrained to be
$10^{10.3}$ $\le $ $n_{\rm H}$  $\le$  $10^{11.4}$ cm$^{-3}$, -1.87 $\le$ log$ U_E$ $\le$ -1.73,
and $N_{\rm H}$ $ \sim $ $10^{21}$ cm$^{-2}$. Here,  it must be  mentioned that   these parameter ranges are derived only from three absorption lines. Due to the lack of  a high-quality UV spectrum, we cannot rule out other possible solutions.
With the ionization parameter, we constrain the radius of the HV BAL gas,
$r$ $\sim$ 0.5pc, comparable to the distance of  the normal BLR.
The LV component corresponds to -3500 to -1000 \kmps\ and  is only detected in \heiozetz\ and \heiteen.
The observations for this component are not sufficient to constrain the characteristic of
 the corresponding BAL outflow. Nevertheless,  a qualitative analysis indicates that the LV BAL gas seems to have a larger $U$ and column density than the HV BAL gas.  Also, the distance  from the LV BAL gas to the central engine is smaller than that of the HV BAL gas
In addition, the analysis  of the BELs of SDSS J0751+1345 indicates that the blueshifted emission components
are detected in \aliii\ and \mgii\ and the corresponding line ratio of \aliii/\mgii\ can be
constrained in the range  of 0.23--0.95. This line ratio can be reproduced by the outflow with the
physical conditions of  the HV BAL outflow, which implies that the BAL and BEL outflows
may be connected in SDSS J0751+1345.

\subsection{Energetic Properties of the Outflow}
According to the discussion in Borguet et al. (2012), for a thin  ($\Delta R/R \ll 1$ ) outflow shell,  with a radius to the central source R,  a radial velocity is v,  a column density  $N_{H}$ and  a global covering fraction $\Omega$,  its  mass-flow rate ($\dot{M}$) and kinetic luminosity ($\dot{E_{k}}$) can be derived through the equations:

 \begin{equation}
 \dot{M}=4 \pi R \Omega \mu m_{p} N_{H} v
 \end{equation}

 \begin{equation}
 \dot{E_{k}}=2 \pi R \Omega \mu m_{p} N_{H} v^3
 \end{equation}

 In the two equations, $m_{p}$ is the mass of  a proton and $\mu $, the mean atomic mass per proton, is equal to 1.4.  Usually,
 $\Omega$ of  the BAL outflow gas is replaced by the fraction of BAL quasars. In optical-selected quasars, the fraction is
about 10\%-20\% (e.g., Trump
et al. 2006;  Liu et al. 2016; Zhang et al. 2017). For the HV BAL outflow,
with R = 0.5 pc, $N_{H}$ = 10$ ^{21} $ cm$ ^{-2} $, radial velocity $v \sim $ 6000 \kmps\
and $\Omega$ $\sim$0.15, we can estimate its $ \dot{M} $  and $ \dot{E_k} $ as $ \dot{M} $ = 0.04 M$ _{\odot} $ yr$ ^{-1} $ and $ \dot{E_k} $ = 7.8 $ \times $ 10$ ^{41} $ erg s$ ^{-1} $, respectively. Compared to the Eddington  luminosity (L$ _{EDD} $), which is about 5$ \times $ 10$^{46} $ erg s$ ^{-1} $,   $ \dot{E_k} $ of the HV BAL outflow is less than 10$ ^{-4} $ L$ _{EDD} $.

According to the previous studies, for a high-velocity AGN outflow that has efficient feedback to the host galaxy, its ratio of $ \dot{E_k} $/L$ _{EDD} $ should be as large as  a few percent (e.g. Scannapieco \& Oh 2004; Hopkins \& Elvis 2010; Zhang et al. 2017). For the HV outflow in our source, the ratio  $ \dot{E_k} $/L$ _{EDD} $ is too low to drive the feedback efficiently.
However,  it should be noted that the  $ \dot{E_k} $ of  the HV BAL outflow is only a  lower limit on the $ \dot{E_k} $ of the outflows in SDSS J0751+1345
because the LV BAL outflow and the BEL outflow are not considered in the estimation above.

\subsection{Other Quasars with both BAL and BEL Outflows}
Blueshifted BALs and BELs provide unique  diagnostics of the physical conditions of outflow gas in quasars.
Some  studies the quasars with both blueshifted \civ\ BEL and BAL simultaneously detected  has suggested that there may be  a relationship between the blueshifted  BEL and BAL outflows. Richards et al. (2011) found that, compared to  normal quasars, the \civ\ BEL in  BAL quasars tend to have larger blueshifted velocities. Rankine et al. (2019)  found that for the quasars with  a fixed \civ\ luminosity, with the increase of  the \civ\ blueshifted velocity, a larger fraction of quasars are detected with \civ\ BAL. However,  these studies mainly focus on the blueshifted BELs and BALS of high-ionization lines. In quasars with high-ionization BALs,  only 15\% show low-ionization BALs. Also, compared to the ubiquitous blueshifted CIV BEL (e.g. Wang et al. 2011; Shin et al. 2017), the low-ionization lines, for example \mgii, are often considered  to have no blueshited  (Marziani et al. 2013, Popovic et al. 2019) and are often used to derive the systematic redshift of quasars (Bonning et al. 2007, Shen et al. 2011).

Besides the \civ\ blueshfited BEL and BAL,
the quasar J1634, which is reported in Liu et al. (2016), is observed with the blueshifted BELs and BALs likely
originating from the same outflow. Different from SDSS J0751+1345, the density of the outflow material
$ n_{\rm H} $ is about 10$ ^{5} $ cm$ ^{-3} $ and the distance to the central BH is about 50 pc,
which indicate that the outflow is  located at the scale of  the narrow line region of quasars.

Liu et al. (2019) reported a quasar, J1633, in whose spectra both low-ionization blueshifted BEL and BAL outflows are detected.
The density   of the blueshifted BEL outflow is n$\rm _{H}$  $\sim
$  10$^{10.6}$-10$^{11.3}$ cm$^{-3}$, column density N$\rm _{H}\geq
10^{23}$ cm$^{-2}$, and ionization parameter $U \sim
10^{-2.1}-10^{-1.5}$. The distance from the central source is similar to the size of  the normal BLR.
The BAL outflow shares the same physical parameters  with the  BEL outflow except for the column density
which is in the range  of $10^{21}$ $\le$ log(N$_{\rm H})$ $\le$ $10^{21.4}$ cm$^{-2}$,
about two orders of magnitude less than that derived for the BEL outflow.
Assuming  that the BAL and BEL  outflows come from the same gas, to explain this N$_{\rm H}$ diffusion ,
we proposed a multicolumn density model for the outflow gas  having an increasing global covering fraction  with a decreasing column density.
In this model, the sight line along which the BALs are produced
will have a high probability to peer through the outflow gas with a higher global covering fraction, which means this outflow gas has a lower column density.
Conversely, the blueshifted BELs  originate more likely from the the  higher column density outflow gas .
In SDSS J0751+1345, we find that the physical parameters for the BAL outflow, such as
$ n_{\rm H} $ and  the ionization parameter U,  are very close to  those of the BEL outflow in J1633.
The $ N_{\rm H} $ of  the BAL outflow in our source is about two orders of magnitude less than
that of  the BEL outflow in J1633 but similar to  that of the value derived from the BAL  in J1633.
Furthermore, the distance  from the BAL outflow to the central engine in SDSS J0751+1345
is also similar to the BEL outflow of J1633, both  located at the scale of  the normal BLR.
The similarity of  the physical properties of the outflow gas in SDSS J0751+1345 and J1633 may suggest
the common origin and  geometrical distribution of  the outflows.
Future simultaneous observations of blueshifted BELs and BALs in the spectra of more quasars
will be crucial  for testing the universality of  the production of the outflows.

\begin{deluxetable}{cccc}
\tabletypesize{\scriptsize}
\tablewidth{0pt}
\tablenum{1}
\tablecaption{Photometric Observations of SDSS J0751+1345}
\tablehead{
\colhead{Wavelength Band/Range}  & Mag. &  Survey  & \textit{MJD}  }
\startdata
\  $\textit{u}$          &$ 19.16\pm 0.02 $ &SDSS     &53352\\
\  $\textit{g}$          &$ 18.42\pm 0.01 $ &SDSS     &53352\\
\  $\textit{r}$          &$ 18.00\pm 0.01 $ &SDSS     &53352\\
\  $\textit{i}$          &$ 17.85\pm 0.01 $ &SDSS     &53352\\
\  $\textit{z}$          &$ 17.50\pm 0.02 $ &SDSS     &53352\\
$  J  $               &$ 16.31\pm 0.13 $ &2MASS   & 50752\\
$  H  $               &$ 15.62\pm 0.15 $ &2MASS   & 50752\\
$  K_{s}  $               &$ 14.95\pm 0.14 $ &2MASS   & 50752\\
$  W1 $               &$ 13.35\pm 0.02 $ &WISE   &55489 \\
$  W2 $               &$ 12.03\pm 0.03 $ &WISE   &55489 \\
$  W3 $               &$ 9.02\pm 0.03 $ &WISE   &55489 \\
$  W4 $               &$ 6.85\pm 0.09  $ &WISE   &55489 \\
\enddata
\end{deluxetable}

\end{document}